\documentclass[twocolumn,showpacs,eqsecnum,preprintnumbers,amsmath,amssymb,aps,prb,floatfix,superscriptaddress]{revtex4-1}

\usepackage{subfigure}
\usepackage{amsfonts}           
\usepackage{amssymb}            
\usepackage{amsmath}            
\usepackage{comment}
\usepackage{latexsym}           
\usepackage{dcolumn}            
\usepackage[dvips]{graphicx}   
\usepackage{enumerate}
\usepackage{epstopdf}           
\usepackage{fancyhdr}           
\usepackage{hyperref}           
\usepackage{color}
\usepackage{setspace}           
\usepackage{xspace}             
\usepackage{subfigure}          
\usepackage{mdframed}
\usepackage{bm}                 
\usepackage[ulem=normalem]{changes}

\newcommand{\rw}{\rightarrow}
\newcommand{\bea}{\begin{eqnarray}}
\newcommand{\eea}{\end{eqnarray}}

\def\YBCO{YBa$_2$Cu$_3$O$_{6+x}$}

\def\BSCCO{Bi$_2$Sr$_2$CaCu$_2$O$_{8+\delta}$}

\def\C60{A$_x$C$_{60}$}

\def\ie{ {\it i.e.\/} }

\def\SROtwo{ Sr$_{3}$Ru$_{2}$O$_{7}$}

\def\HgCu3{HgCa$_2$Cu$_3$O$_{8+y}$}
\def\HgCu4{HgBa$_2$Ca$_3$Cu$_4$O$_{10+y}$}
\def\TlCu{Tl$_2$Ba$_2$CuO$_{6+\delta}$}
\def\TlCu3{Tl$_2$Ba$_2$Ca$_2$Cu$_3$O$_{10+y}$}
\def\TlCu4{Tl$_2$Ba$_2$Ca$_3$Cu$_4$O$_{12+y}$}

\def\BiCu3{Bi$_2$Sr$_2$Ca$_{2}$Cu$_3$O$_y$}

\def\8LSCO{La$_{1.88}$Sr$_{.12}$CuO$_4$}
\def\110LNSCO{La$_{1.5}$Nd$_{0.4}$Sr$_{0.1}$CuO$_{4}$}
\def\stage4LCO{La$_{2}$CuO$_{4+\delta}$}
\def\Y248{YBa$_2$Cu$_4$O$_8$}

\def\NbSe2{NbSe$_2$}
\def\TaSe2{TaSe$_2$}
\def\TiSe2{TiSe$_2$}
\def\NaCoOH2O{Na$_{0.3}$CoO$_{2y}$H$_2$O}
\def\MgB2{MgB${}_2$}

\def\URu2Si2{URu$_2$Si$_2$}
\def\LFAO{LaFeAsO}
\def\Ba122{Ba(Fe$_{1-x}$Co$_x$)$_2$As$_2$}

\newcommand{\cf}{{\it cf.~}}

\def\n{\mathbf{n}}
 \def\r{\mathbf{r}}
 \def\C{\mathcal{C}}
 \newcommand{\ud}{\textrm{d}}

\begin{document}

\title{Ising nematic fluid phase of hard-core dimers on the square lattice}

\author{Stefanos Papanikolaou}
\affiliation{Departments of Mechanical Engineering \& Materials Science, Yale University, New Haven, Connecticut 06520}
\affiliation{Department of Physics, Yale University, New Haven, Connecticut 06520}

\author{Daniel Charrier}
\affiliation{Max-Planck-Institut f{\"u}r Physik komplexer Systeme, N{\"o}thnitzer Str. 38, 01187 Dresden, Germany}

\author{Eduardo Fradkin}
\affiliation{Department of Physics and Institute for Condensed Matter Theory, University of Illinois at Urbana-Champaign, 
1110 West Green Street, Urbana, Illinois 61801-3080}

\date{\today}
\begin{abstract}
We present a model of 
classical 
hard-core dimers on the square lattice that contains an Ising nematic phase in its phase diagram. 
We consider a model with an attractive interaction for parallel dimers on a given plaquette of the square lattice 
and an attractive interaction for neighboring parallel dimers on the same  row ({\it viz} column) of the lattice. 
By extensive Monte carlo simulations we find that with a finite density of holes  the phase diagram has, with rising temperatures,  
a columnar crystalline phase, an Ising nematic liquid phase and a disordered fluid phase,
separated by Ising continuous phase transitions. We present strong evidence for the Ising universality class of both transitions.
The Ising nematic phase can be interpreted as either 
an intermediate classical thermodynamic phase (possibly of a strongly correlated antiferromagnet) or as a phase of a 2D 
quantum dimer model using the Rokhsar-Kivelson construction of exactly solvable quantum Hamiltonians.
\end{abstract}

\maketitle 

\section{Introduction}

Hard-core dimer models have been useful in identifying and handling basic aspects of the configurational frustration in 
quantum spin systems with non-magnetic ground states such as RVB wavefunctions and valence bond singlet crystals.  
Hard-core dimer models have been used extensively in the context of quantum condensed matter model constructions, 
such as the quantum dimer models by Rokhsar and Kivelson,\cite{Rokhsar:1988vn} 
to characterize topological fluid quantum  phases,\cite{Moessner:2001ys} strongly correlated 
superconducting phases,\cite{Fradkin:1990uq} as well as crystalline phases and  stripe phases\cite{Papanikolaou:2007qf} 
(see, e.g Ref.[\onlinecite{Fradkin-2013}]). 
\begin{figure}[b]
 \centering
 \includegraphics[width=0.3\textwidth]{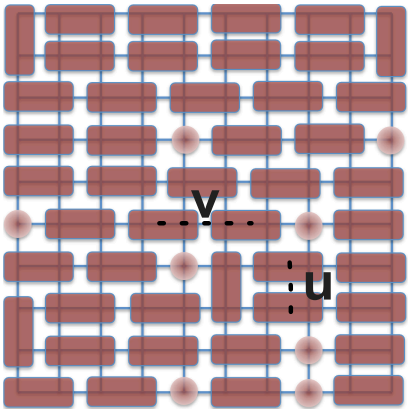}
  \caption{A typical configuration of an Ising nematic phase for a hard-core dimer model with a finite density of holes (see text).}
 \label{fig:1}
 \end{figure}
Such modeling led to useful insights and basic intuitive understanding of 
aspects of quantum frustration.\cite{Moessner-2000,Moessner-2001b} 
Similar models can be used to understand the high-temperature behavior of such strongly correlated 
quantum antiferromagnets, assuming that the valence bond spin gap has an onset at much higher energies.

 \begin{figure*}[hbt]
 \centering
\includegraphics[width=\textwidth]{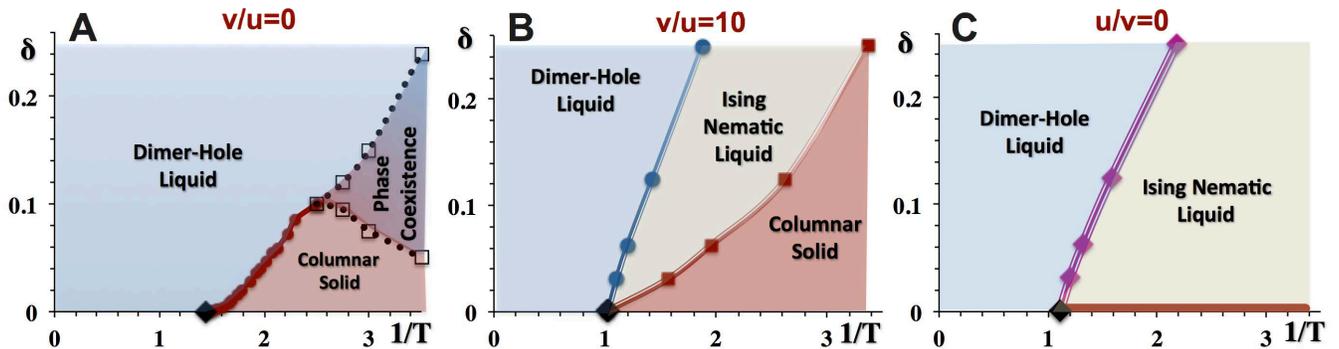} 
 \caption{
 Evolution of phase diagram for different hole densities $\delta$, and for different attractive dimer plaquette interactions $u$ and 
 attractive dimer-aligning interactions $v$ as a function of temperature $T$.
A) When only the plaquette interactions are present there is
a single phase transition is present between a featureless dimer-hole liquid and the columnar solid.\cite{Papanikolaou:2007qf}  
At zero hole density this transition is in the KT universality class.\cite{Alet-2005} At finite hole density there is a line of varying critical 
exponents lead to a rather unconventional multicritical point 
 at $\delta_{m}=0.11,$ $T_m=0.4$ (endpoint of solid line). At lower temperatures $T<T_m$ the transition is discontinuous, and when in a fixed-density ensemble there is phase coexistence. The dotted lines represent the range of observed coexistence, derived from simulations.\cite{Papanikolaou:2007qf}
B) As soon as the dimer-aligning interaction $v$ is finite, a new phase emerges in-between the dimer-hole liquid and the columnar solid, the 
Ising nematic dimer liquid. The phase transition from the columnar solid to the nematic liquid appears to resemble quantitatively (almost 
overlapping) the columnar solid to dimer-hole liquid transition line at zero nematicity. The transition line is  in the Ising universality class. At 
higher hole densities there is a subsequent Ising transition line separating the Ising nematic phase from the isotropic dimer-hole liquid. 
C)  When only the dimer-aligning interaction $v$ is present, the columnar solid phase is absent. Only a single Ising transition line separates 
the nematic liquid from the dimer-hole liquid. In the close-packed limit $\delta=0$ and for $u=0$, the nematic liquid is infinitesimally unstable to 
a columnar solid, possibly at finite tilt,\cite{Fradkin:2004fk,Vishwanath-2004} due to a sub-extensive entropy of aligned dimer lines. 
The solid line for $T>1$ signifies the presence of the finite tilt columnar solid. 
Due to the presence of this sub-extensive entropy, the nature of the transition, expected to be in the KT universality class, is unclear from the simulations.
}
 \label{fig:2}
 \end{figure*}

Ising nematic liquid phases and more generally, electronic liquid crystalline phases,\cite{Kivelson:1998uq} have been observed in several 
physical systems, notably in the two-dimensional electron gas (2DEG) in large magnetic fields, in the bilayer ruthenate {\SROtwo} in magnetic 
fields, in the iron-based superconductors and most interestingly, and in the cuprate 
superconductors.\cite{Fradkin-2010,Fradkin:2012kx,Kivelson:2003vn} 
In particular, nematic order has been reported in {\YBCO} in much of the pseudogap regime,\cite{taillefer-nematic-2009} 
and in very underdoped {\YBCO} over a large temperature range,\cite{hinkov-2007} as well as in {\BSCCO},\cite{lawler-2009}
and  in the iron-based superconductors {\LFAO} and {\Ba122}.{\cite{delacruz-2008,chu-2010,chuang-2010,Fang-2008,Xu-2008}

In this paper, we study the emergence of an Ising nematic dimer phase in a  classical dimer model, using extensive numerical 
classical Monte Carlo simulations and analytical arguments.
We consider dimer models with two types of interactions: a) an attractive interaction with strength $u$ for parallel dimers 
on the same plaquette of the square lattice (which favors columnar ordered states), and b) an attractive interaction of strength $v$ 
for dimers on next-nearest-neighbor bonds on the same  row ({\it viz} column)  of the square lattice (which favors all tilted columnar states). 
We investigated the phase diagram of this model, shown schematically in Fig.~\ref{fig:2} A, B and C,  at full dimer coverage and at a finite 
density $\delta$ of holes as a function of temperature $T$ and for varying dimer interactions $u$ and $v$.

In the fully packed regime, 2D dimer models can describe ordered phases, columnar phases  with varying degree of translation 
symmetry breaking,\cite{Fradkin:2004fk,Vishwanath-2004,Alet:2006zr,Castelnovo-2006,Papanikolaou:2007qf} 
as well as quantum critical systems\cite{Rokhsar:1988vn,Ardonne:2004nx} and topological phases.\cite{Moessner:2001ys,Fendley-2002}  
The nematic phase of this system is characterized by spatially disordered dimer configurations that globally break rotational invariance, 
as shown in Fig.~\ref{fig:1}, i.e. long range orientational order that breaks spontaneously the $C_4$ point group symmetry of 
the square lattice but leaves translation symmetry intact. We observe that such a phase cannot be stable on the square lattice 
against columnar energy perturbations in the absence of holes (the undoped system). Only in the existence of mobile holes, 
we find that a nematic phase exists for a finite temperature range (see Fig.~\ref{fig:2}B). 
Moreover, we find that in the presence of generic perturbations such as a small attractive columnar interaction, 
this liquid phase only occurs in an intermediate temperature range, comprised between a high temperature isotropic liquid 
and a low temperature columnar crystal, separated by corresponding (classical) 2D Ising phase transitions.

These phases can be interpreted as occurring in a classical dimer model at various coverings.  
In the context of quantum antiferromagnets, a bond occupied by a dimer means that it is  occupied by a valence bond singlet of 
 two neighboring  spins\cite{Anderson-1987,Kivelson-1987} or as a label of an unsatisfied bond in 
 frustrated quantum magnets.\cite{Moessner-2001b,Bergman-2007} In doped antiferromagnets with a spin gap, 
 in addition to valence bonds one has to include the charge degrees of freedom in the form of holes. 
 A simple model of such systems is the doped quantum dimer model.\cite{Rokhsar:1988vn,Fradkin:1990uq}
  Dimer models of the type we discuss here can also serve to represent, qualitatively, forms of orbital  
 order of the type used in models of the iron superconductors,\cite{Lv-2011} 
 and in the analysis of the STM data in {\BSCCO},\cite{lawler-2009} 
 as well as for the analysis on valley order in certain 2DEGs in magnetic fields.\cite{Abanin-2010,Kumar-2013} 
 Although the dimer models do not represent many aspects of the microscopic physics of these systems faithfully,
  its symmetries are the same as the point groups of the more microscopic models and, 
  hence, should be useful to describe many aspects of these systems, particularly their phase transitions.

Here we show that a classical dimer model with a finite density of holes can lead, with suitable interactions, 
to a nematic phase in dimer models arising from the 
melting of a columnar state.
However, it has been unknown how to stabilize such a nematic liquid in classical or/and quantum dimer models by using 
isotropic Hamiltonians with local interactions. 
By using the RK-construction,\cite{Rokhsar:1988vn,Ardonne:2004nx,Castelnovo-2005,Papanikolaou:2007qf} 
it is straightforward to find an exactly solvable quantum dimer models  with a stable quantum nematic phase whose ground-state 
wave function $\big| \Psi \rangle$ has local weights of the same form as the Gibbs weights of the classical dimer model we study here, 
given the configurations are assumed to be orthonormal. Namely,
\begin{equation}
\big| \Psi \rangle=\frac{1}{\sqrt{Z}} \exp(-\beta E_{\mathcal{C}}/2) \big| \mathcal{C}\rangle
\end{equation}
where 
\begin{equation}
Z = \sum_{\{\mathcal{C}\}} \exp(-\beta E_\C)
\label{eq:PF}
\end{equation}
is the norm of the wave function and also the partition function of a classical system with Gibbs weight 
$\exp(-\beta E_{\mathcal{C}})/Z$ in the same dimension.

We fully characterize the behavior of the corresponding classical model  
through analytic arguments and extensive simulations. First, we show that the key ingredient for stabilizing nematic dimer liquids is 
a strong same-row  or same-column dimer-aligning, spatially isotropic,  interaction. In a fully packed lattice this interaction leads to 
crystalline phases.\cite{Alet:2005cr, Charrier:2008ve, Charrier:2010ly, Alet:2006zr, Papanikolaou:2007qf} Here we will show that at finite hole 
density it stabilizes  an Ising nematic fluid phase. 
Further, we demonstrate numerically that all relevant transitions at finite hole density are of Ising nature and that the nematic phase, 
even though it breaks explicitly rotational symmetry, by having most dimers point to the same direction, 
also has disordered liquid-like correlations along the orientation direction. 
However, in the close-packed case, and when only the dimer-aligning interaction is present, it is unclear whether the transition is in the Ising or  the KT universality class. Theoretical arguments predict the KT transition but the simulations give, as we will discuss, inconsistent results possibly due to finite-size effects.

This work is organized as follows. In Sec. \ref{sec:model} we present the generalized dimer model and discuss its symmetry properties. 
In Sec. \ref{sec:MC} we present and analyze the results of our Monte Carlo simulations which lead to the phase diagrams of Fig.~\ref{fig:2}. 
In Sec.~\ref{sec:MFT} we present results form a mean-field theory based on a formulation of the generalized dimer model in a 
Grassmann variables formulation. Our conclusions are presented in Sec.~\ref{sec:conclusions}.

\section{The Model}
\label{sec:model}

The classical dimer model is on a square lattice of linear dimension $L$ (total number of sites $N=L^2$) covered by 
$N_d$ hard-core dimers and $N_h=L^2-2N_d$ holes with the hole density being $\delta=N_h/L^2$
The configuration space consists of all the possible dimer coverings satisfying the hard-core constraint with a given number of holes. 
The energy of a dimer configuration $\cal C$ is
\begin{align}
E_{\cal C} &=\label{eq:EC}\\
- &\sum_{{\bf r}, \alpha = x,y}  \left[ v\; n_\alpha({\bf r})n_\alpha({\bf r}+2 \hat{\bf e}_\alpha) + u\; n_\alpha({\bf r}) n_\alpha({\bf r}
+ \hat{\bf e}_{\beta\neq\alpha}) \right]\nonumber
  \end{align}
where $n_\alpha({\bf r})$ is the dimer occupation number of the link of the square lattice with endpoints at the two nearest neighbor sites 
$\bf r$ and ${\bf r}+ {\hat {\bf e}}_\alpha$ (with $\alpha=x,y$).
The first term of the total energy $E_{\cal C}$ of Eq.~\eqref{eq:EC} is proportional to the number of parallel dimers on the same row 
(or column) on next-nearest-neighbor sites, while the second term is proportional to the number of plaquettes with two parallel dimers,
 $\beta=1/T$ is the inverse  temperature of the classical model and we assume that the interaction of parallel dimers on 
a plaquette is attractive, $u>0$. Below, we will refer to the first term of Eq.\eqref{eq:EC}  the {\it dimer-aligning interaction} 
(with strength $v$) and to the second term  the {\it plaquette interaction} (with strength $u$).

The model with configurational energy $E_{\mathcal{C}}$ of Eq.~\eqref{eq:EC} is explicitly invariant under translations 
(by one lattice spacing) in both directions of the square lattice and under $C_4$ point group symmetries. 
However, in the special case $u=0$, the configurational energy is also invariant under the independent rigid shifts of the dimer 
configurations on the links of each separate rows and columns. This discrete ``sliding'' symmetry formally relates dimer 
configurations with different tilts.\cite{Fradkin:2004fk} As it stands, this symmetry formally leads to  a sub-extensive 
(on the square lattice) entropy even at $T=0$ which hinders the equilibration of this system for large lattice sizes. 
A discrete sliding symmetry of this type also arises in the 2D classical Ising model with only four spin interactions.\cite{Xu-2004} 
However this symmetry can be broken by fixed  boundary conditions to select a particular tilted state. 
This symmetry is also explicitly broken by the plaquette interactions even for arbitrarily small values of $u$. 
For this reason, it is important to focus on the case $u>0$.

The partition function for a dimer model with no aligning interaction $v=0$ was studied in detail both in the fully packed 
case\cite{Alet:2005cr,Alet:2006zr,Papanikolaou:2007qf} and with a finite density of holes.\cite{Papanikolaou:2007qf} 
In the fully packed regime it was found that, for $u\geq 0$, the system is  characterized by a line of fixed points parametrized by $u=1/T$. 
At temperatures below a Kosterlitz-Thouless (KT) transition at $ u_c\simeq 1.5$, the fully-packed dimer model enters  
a four-fold degenerate columnar phase which has a spontaneously broken translation and rotational 
symmetry.\cite{Alet:2005cr,Papanikolaou:2007qf} 
For any density of holes, the line of fixed points is replaced by a disordered phase, 
but the columnar phase survives, unless the density of holes is large enough.\cite{Papanikolaou:2007qf}

The phase diagram in the presence of the aligning interaction $v$ as well as the plaquette interaction $u$ at various hole densities is complex. Apart from the already described disordered liquid and columnar solid phases, it is expected that a new phase emerges, which only breaks the rotational symmetry but not the translational one. The existence of this Ising nematic phase has been proven for the case $u=0$ on dilute systems by Heilmann and Lieb~\cite{heilmann-79}. Here, we will explore the emergence of this phase primarily through numerical simulations, supplemented by theoretical arguments. 

Intuition for this model can come from the fact that the close-packed square lattice dimer model has an effective 
field theoretic description in terms of a continuum Gaussian (free boson) field. 
The mapping of the square lattice classical dimer model to a height model,\cite{Ardonne:2004nx} 
proceeds by assigning a height variable on each plaquette. On the even sublattice, going clockwise, 
the height changes by $+3$ if a dimer is on a link, and by $-1$ when there isn't one. 
On the odd sublattice, the changes reverse. 
Moreover, the dimer hard-core constraint implies that there are only $4$ local dimers configurations on each site, 
guaranteed in the dual height model by the compactification $h\rw h+4$. In terms of the rescaled field $\phi=\pi h/2$ the actual action is
\begin{equation}
S=\int d^2x \left[ \frac{K}{2} \left(\nabla \phi\right)^2+g \cos(4\phi)\right]
\label{Seff-0density}
\end{equation}
with $K=1/4\pi$ at the non-interacting dimer point and with a charge $4$ perturbation, $\cos
(4\phi)=\cos(2\pi h)$, biasing the coarse-grained height field to take integer values. 

The different observables (and hence perturbations) of this theory consist of the $O_{n,m}$ composite operators with $n$ units of charge and $m$ units of vorticity.\cite{Kadanoff-1979,Kadanoff-1979b} Their scaling dimensions are 
\begin{equation}
\Delta(n,m)(K)=\frac{n^2}{4\pi K} +\pi K m^2
\end{equation}
As shown in Ref.[\onlinecite{Papanikolaou:2007qf}],
the columnar order parameter corresponds to the elementary charge operator $O_{\pm 1, 0}$, which is relevant with $\Delta_{1,0}=1$, while the nematic order parameter corresponds to $O_{\pm2,0}$ which is irrelevant at the free dimer point. The operator $O_{\pm 4, 0}$, which naturally appears due to discreteness of the height values is a strongly irrelevant operator, but it is the one that drives the KT transition, when the plaquette attractive interaction is added. Moreover, the hole operator, corresponding to a hole density, is represented by the fundamental vortex operator $O_{0,\pm1}$, a typically strongly relevant operator which drives the system to a high temperature liquid. Finally, by using OPE based on the density operator definitions~\cite{Fradkin:2004fk}, it is easy to show that the leading effect of the dimer-aligning interaction on the same row or the same column at the non-interacting dimer point is just a renormalization of the free field stiffness. This conclusion is verified in the simulations, since in the close-packed limit (and $u$ infinitesimal) the columnar crystalline phase appears. On the dilute lattice, such a picture breaks down and the emergence of the phase diagram is not understood analytically. 

\section{Results from Monte Carlo Simulations}
\label{sec:MC}

 We simulate the partition function of  this dimer model with configurational energy $E_{\mathcal{C}}$ of Eq.\eqref{eq:EC}  by means of a  Monte Carlo worm-algorithm with a local heat-bath detailed balance condition.\cite{Alet:2006zr} We have set the energy scale to be $v=1>0$.
 The study is made in the canonical ensemble, similar to previous analogous studies.\cite{Papanikolaou:2007qf, Alet:2006zr,Charrier:2008ve,Charrier:2010ly} The observables, apart from the energy related specific heat, are related to the possible symmetry breaking, the columnar and nematic order.
The local columnar order parameter $\mathbf{c}(\r)$~\cite{Alet:2005cr,Papanikolaou:2007qf} is defined with respect to the dimer 
occupation number at each site 
$n_\alpha(\r)$, 
\bea
\mathbf{c}(\r) = \sum_{\alpha = x,y} (-1)^{r_\alpha} n_\alpha(\r)\hat{e}_{\alpha}
\eea
 with the global order parameter being 
 \begin{equation}
 \mathbf{C} = \frac{2}{L^2} \sum_\r \mathbf{c}(\r)
 \end{equation}
  and is normalized such that the four columnar states correspond to
  \begin{equation}
  \mathbf{C} = \{ \pm 1, 0 \} , \{ 0 , \pm 1\}.
  \end{equation}
   A columnar state breaks the rotation and the translation symmetries of the lattice.

 The rotation symmetry breaking (nematic) order parameter is defined to be 
 \begin{equation}
 R =  \frac{2}{L^2} \left| N_\C^h - N_\C^v \right|
 \end{equation}
  where $N_\C^h$ and  $N_\C^v$ are, respectively, the number dimers on horizontal and  vertical bonds in configuration $\C$.
 One also considers the corresponding susceptibilities $\chi$. In particular, for a second order transition 
 \begin{equation}
 \frac{\chi}{N} = \frac{\langle R^2\rangle - \langle |R|\rangle^2}{L^3} \propto L^{2-\eta}
 \end{equation}
 while for a first order transition 
 \begin{equation}
\frac{ \chi}{N} \propto L^2
 \end{equation}
  Finally, the Binder cumulant
  \begin{equation}
  B = \frac{\langle R^4\rangle}{\langle R^2\rangle^2}
  \end{equation}
  is a scale-invariant quantity in the case of a continuous transition, and should thus exhibit a crossing point at criticality as a function of the system size. In addition, the finite size-scaling (FSS) of its derivative with respect to the temperature, 
  \begin{equation}
  \frac{\ud B}{\ud T} \propto L^{1/\nu}
  \end{equation}
   provides a direct way to determine the critical exponent $\nu$ of the correlation length for nematic fluctuations.

\subsection{Case $u=0$, $\delta=0$:  attractive dimer-aligning interactions (but no plaquette interactions) on a close-packed dimer lattice}
\label{ssec:MC-1}

\begin{figure}[t]
\centering
 \includegraphics[width=0.4\textwidth]{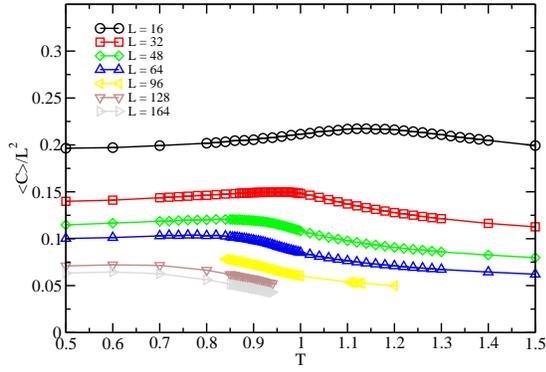}
 \caption{Columnar order parameter as a function of temperature for multiple system sizes in a close-packed dimer  model without a plaquette interaction, $u=0$ and zero hole density, $\delta=0$. The columnar order vanishes quickly in the thermodynamic limit, with no signs of an instability.}
 \label{fig:MC1a}
 \end{figure}

\begin{figure}[b]
\centering
\vspace{0.2cm}
 \includegraphics[width=0.4\textwidth]{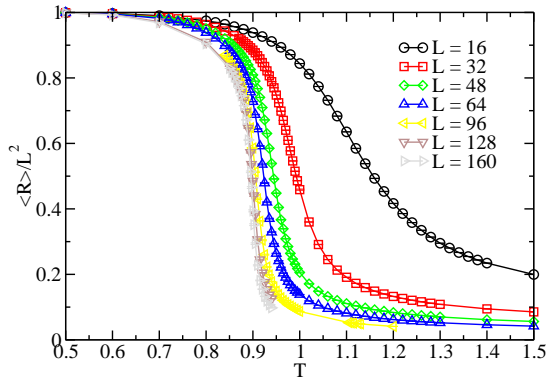}
 \caption{ Rotational order parameter as a function of temperature for multiple system sizes in a close-packed dimer  model without a plaquette interaction, $u=0$ and zero hole density, $\delta=0$.  Rotational order displays a clear transition.}
 \label{fig:MC1b}
 \end{figure}

\begin{figure}[hbt]
\centering
 \includegraphics[width=0.4\textwidth]{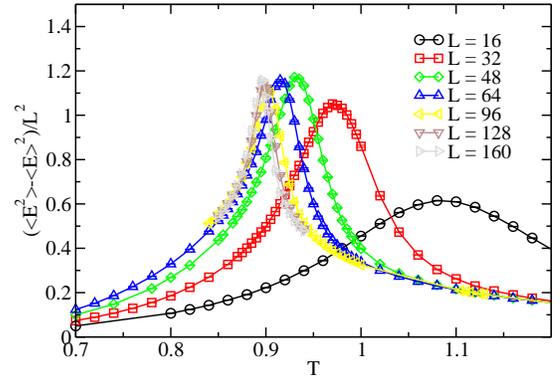}
 \caption{  Specific Heat for different system sizes in a close-packed dimer  model without a plaquette interaction, $u=0$ and zero hole density, $\delta=0$. The specific heat clearly displays a cusp or possibly a logarithmic divergence, consistent with both an Ising 
 and a KT transition. }
 \label{fig:MC1c}
 \end{figure}

\begin{figure}[hbt]
\vspace{1cm}
\centering
 \includegraphics[width=0.4\textwidth]{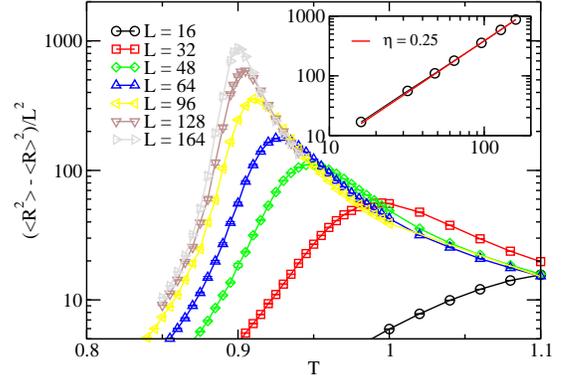}
 \caption{Rotational order parameter susceptibility displaying a divergence with an anomalous exponent $\eta=0.25$ consistent 
 with an Ising transition, in a close-packed dimer  model without a plaquette interaction, $u=0$ and zero hole density, 
 $\delta=0$. The position of the peak appears to be very close to the position of the specific heat peak. 
 For a KT transition, there would be expected a discrepancy of the two peaks.}
 \label{fig:MC1d}
 \end{figure}

Firstly, we investigate the simplest case, being the case where only the dimer-aligning attractive interaction is present,
$u=0$ and $v=1$, on a close-packed dimer lattice. In this limit the system has a discrete ``sliding'' symmetry which, for 
periodic boundary conditions leads to a sub-extensive ground state entropy. This feature indicates that the ground state may be any of the 
possible tilted phases, such as those discussed in Ref.[\onlinecite{Fradkin:2004fk}]. This feature poses serious numerical difficulties. 
Nevertheless, as we will see in Section \ref{ssec:MC-2}, these difficulties are suppressed once $u>0$, which favors columnar order, no matter 
how small it is. Since in the case of $v=0$ and $u>0$ the transition from the four-old degenerate columnar order state is in the KT 
universality class,\cite{Alet-2005,Papanikolaou:2007qf} here we might expect the same scaling. 
However the existence of many tilted ground states (each with a four-fold global degeneracy) makes the analysis more difficult 
and convergence problematic.

Close-packed dimer configurations on the square lattice present critical correlations at 
$T=\infty$ as aforementioned. 
While from the field theory (and renormalization group arguments) we expect, as described, a renormalization of the 
KT-transition temperature toward a solid phase,which should resemble a columnar solid under appropriate boundary conditions on 
an infinite lattice, our simulations present a 
complex picture shown in Figs.~\ref{fig:MC1a},  \ref{fig:MC1b}, \ref{fig:MC1c}, and \ref{fig:MC1d}. 
From this data, it is reasonable to conclude that this is an Ising transition, given that the anomalous exponent 
of correlation function of the rotational order parameter (having a parent Ising symmetry) appears to be $\eta=0.25$, 
consistent both with KT and with Ising universality, 
and that the columnar order is absent. On the other hand, the specific heat does seem to diverge is consistent with either a cusp 
or with a weak a logarithmic divergence, 
both of which have an exponent  $\alpha=0$. this may be consistent with an Ising transition, which has its prototypical logarithmic divergence.
 In a KT transition, the specific heat displays a peak clearly above the critical temperature (for example detected through the peak of the rotational susceptibility) and does not diverge; In our case, the specific heat is clearly non-divergent but the peak appears unconventionally close ($\pm0.006$) to the critical temperature for the system sizes considered (as noticed in a comparison of Figs.~\ref{fig:MC1c} and \ref{fig:MC1d}). 
 In addition,  Binder cumulants (not shown) 
have non-Ising behavior which display a crossing but not at the 
universal expectations for Ising systems. It may well be that this discrepancy implies that the Ising signatures mask the true nature 
of the transition. The simulations were performed on  double-periodic boundary conditions for reasons of numerical efficiency. 

However, in this 
situation the sub-extensive entropy discussed above is not suppressed. Hence our simulations cannot distinguish a columnar state from any 
state with broken translational invariance but at finite tilt. Since the entropy is 
sub-extensive the associated discrete ``sliding'' symmetry must be spontaneously broken. However,
performing fixed boundary condition simulations  appears numerically intractable. 
For all these reasons we believe that our simulations with periodic boundary conditions have not converged at low temperatures for $u=0$.\\

\subsection{Case $u>0,v>0$, $\delta=0$: Competition between attractive dimer-aligning  
and plaquette interactions on a close-packed dimer lattice}
\label{ssec:MC-2}

\begin{figure}[hbt]
\centering
 \includegraphics[width=0.4\textwidth]{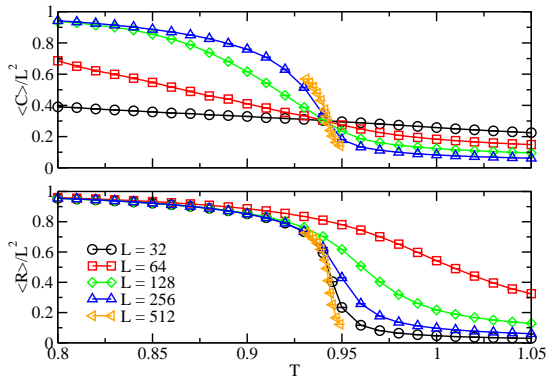}
 \caption{Columnar (top) and nematic (bottom) order parameters in a close-packed dimer model with dimer-aligning and 
 plaquette interactions at zero hole density. The plaquette interaction is  interaction has  $u=0.1$ 
 and the dimer-aligning interactions have  $v=1.0$. 
 In the thermodynamic limit, both order parameters appear to display a strong transition, signifying the emergence of a columnar solid phase. 
 The same happens at any $u$ we studied $(0.2, 0.5)$ (not shown).}
 \label{fig:MC2a}
 \end{figure}

\begin{figure}[bt]
\centering
\vspace{0.5cm}
 \includegraphics[width=0.4\textwidth]{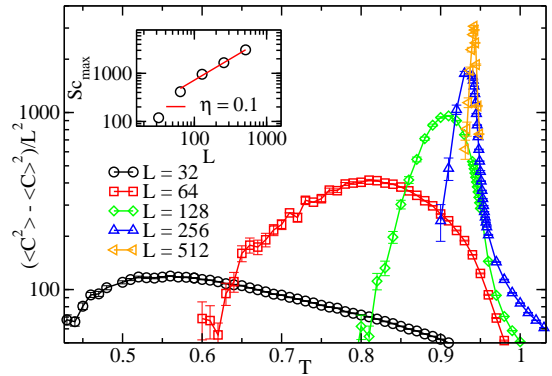}
\caption{Columnar susceptibility for different sizes for a close-packed dimer model with dimer-aligning  
and plaquette interactions and zero hole density.  The dimer-aligning interaction has  $v=1.0$ and the plaquette interaction $u=0.1$. 
Slowly, the susceptibility displays a divergence, but the anomalous exponent we uncovered is very low $0.1$, 
compared to the KT expectation of $0.25$. We believe that this is still a complex crossover due to the competition 
of the columnar solid with the nematic liquid in finite systems, which has a finite entropy, albeit sub-extensive.}
 \label{fig:MC2b}
 \end{figure}

\begin{figure}[bt]
\centering
  \includegraphics[width=0.4\textwidth]{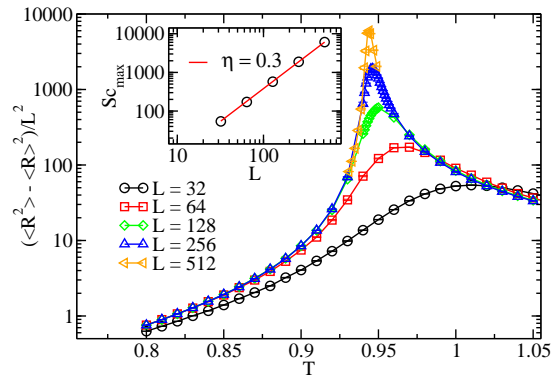}
 \caption{
 Nematic susceptibility in a close-packed dimer model with dimer-aligning and plaquette interactions and zero hole density for
 $u=0.1$ and $v=1.0$.
 Rotational susceptibility for different system sizes. This susceptibility displays a stronger behavior, and an apparent anomalous 
 exponent of $0.3$, close but different from the Ising value. It appears that this exponent also is the outcome of a strong crossover 
 towards the KT universality class, albeit impossible to resolve in simulations for systems up to $512^2$.}
 \label{fig:MC2c}
 \end{figure}

\begin{figure}[bt]
\centering
\vspace{0.5cm}
 \includegraphics[width=0.4\textwidth]{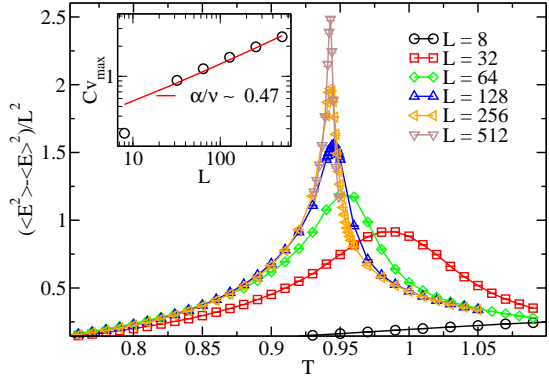}
 \caption{
 Specific heat for different system sizes displaying a divergence with an exponent $\alpha/\nu\simeq0.5$  
 for a close-packed dimer model with dimer-aligning, and plaquette interactions at zero hole density.  Here $v=1.0$ and $u=0.1$. }
 \label{fig:MC2d}
 \end{figure}

\begin{figure}[hbt]
\centering
 \includegraphics[width=0.4\textwidth]{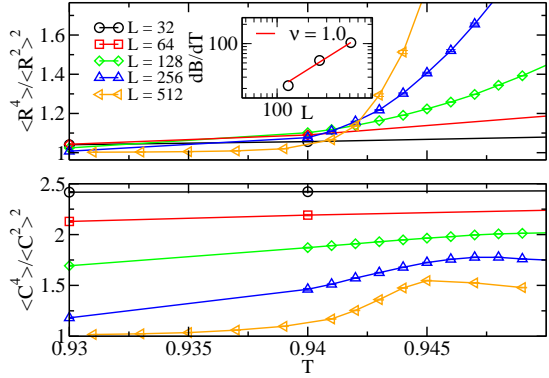}
 \caption{Binder cumulants for the  columnar (top) and nematic (bottom) order parameters in a 
close-packed dimer model with dimer-aligning interactions and plaquette interactions, at zero hole density.  Here, $u=0.1$ and $v=1.0$. 
The columnar order displays no crossing whatsoever, signature of a non-trivial crossover, while the  nematic  order displays a crossing 
at a value that is not consistent with the Ising expectation,  but nevertheless shows the emergence of a correlation length critical exponent 
$\nu\simeq1$ as expected for Ising transitions. These discrepancies suggest that there is a strong crossover even 
at the largest sizes we studied. }
 \label{fig:MC2e}
 \end{figure}

The next case study is for $u$ and $v$ present in the system for a close-packed lattice, but with $u=0.1v$; since we aim to focus on the non-perturbative effects of the dimer-aligning interaction. It is clear that in the opposite limit $v\ll u$ the dimer-aligning perturbation would only shift the location of the KT transition with no change in the universality class. However, in the limit we probe $u\ll v$, the system's behavior is rather complex, exactly due to the non-trivial crossover that we find in the case where $u=0$. While the maximum system size we studied is many times larger than the sizes studied in Ref. [\onlinecite{Alet-2005}], it appears still that the system has not converged to a well defined transition. The expectation is that the system should display a KT transition since the sub-extensive entropic degeneracy is lifted for any finite value of $u\neq 0$. 
In Figs.~\ref{fig:MC2a},  \ref{fig:MC2b}, \ref{fig:MC2c}, \ref{fig:MC2d}, and \ref{fig:MC2e} we present strong evidence that this transition has not converged. The rotational order parameter displays a clear transition with its susceptibility  displaying a large peak with scaling similar to Ising transitions ($\eta=0.25$), and its Binder cumulant shows a crossing point close to the one expected for Ising transitions (1.13), even though there is an apparent ongoing drift of the crossing point toward higher values. The correlation length exponent $\nu$, derived from the slope of the cumulant at the crossing, appears to    be close to the Ising value $\nu=1$, but a drift is observed with increasing system size. But, while these observations appear to be consistent with an Ising transition (similar observations with the $u=0$ case), the specific heat displays a clear transition, with an apparent non-Ising critical exponent $\alpha/\nu\simeq0.5$, while there is some drift as well toward lower values. Further, the columnar order parameter apparently displays a a very weakly-converging transition, where for $L=32$ it is not even detectable (\cf Fig.~\ref{fig:MC2a}). Its susceptibility displays a very weak convergence, with a very wide hump for moderate system sizes, and apparently $\eta\simeq0.1$, but again the shape of the susceptibility does not allow us to conclude that this exponent (derived from just the peak's scaling) can be accurate -- given that in such estimates there is always the implicit assumption of a scaling collapse, \ie the shapes of the curves shall be self-similar in the scaling regime.  Further, its Binder cumulant does not display a crossing, another evidence of a strong crossover. We should note that the KT transition is the only known transition where the Binder cumulants of the order parameter  do not display a crossing (for XY models), so the observed behavior could signify the onset of KT transition scaling. This collection of results suggests that our data is still inside a  strong crossover region even at the largest sizes we studied.

\subsection{Case $u=0$, $\delta>0$: Only dimer-aligning interaction on a dilute dimer lattice and the emergence of an Ising phase 
transition towards a nematic fluid}
\label{ssec:MC-3}

Then, we proceed by considering the effect of finite hole density $\delta$. 
From previous work of some of us\cite{Papanikolaou:2007qf} (shown in Fig.~\ref{fig:2}A), 
there is concrete intuition on the possible expectations when a plaquette interaction is included. 
However, it has not been known how a dimer-aligning interaction affects the phase structure. 
As shown in Fig.~\ref{fig:2}C, we find a new phase, the dimer Ising Nematic phase, which is characterized by orientational order 
of the dimers (as seen in Fig.~\ref{fig:1}) but no translational symmetry breaking, \ie the system remains a fluid. 
It is characteristic that the stability of this phase over possible effects from plaquette interactions is guaranteed by the presence of holes, 
since they provide a source of extensive entropy to the liquid, a feature absent at close packing due to the hard-core dimer frustration. 

\begin{figure}[t!]
\centering
 \includegraphics[width=0.45\textwidth]{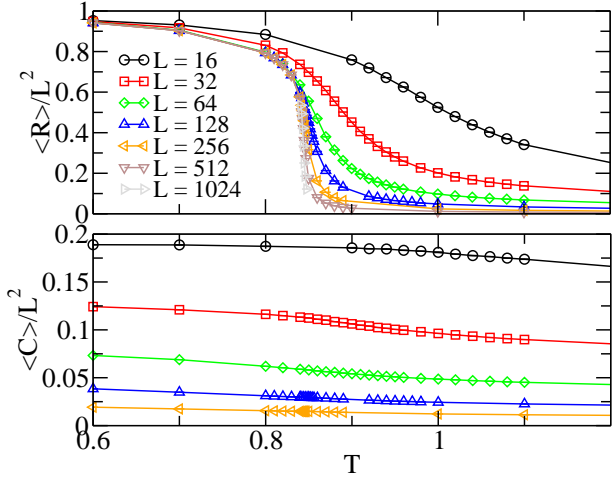}
  \caption{Nematic (top) and columnar (bottom) order parameters in a 
dimer model with only dimer-aligning   interactions. The plaquette interaction is $u=0$ and the dimer-aligning interactions $ v=1$. 
The hole  density is $\delta=0.03125=1/32$.
Top:  the nematic order parameter as a function of temperature for increasing system size, displaying a strong transition. 
Bottom: the columnar order parameter, clearly vanishing in the thermodynamic limit. 
}
 \label{fig:MC3a}
 \end{figure}

\begin{figure}[hbt]
\centering
 \includegraphics[width=0.45\textwidth]{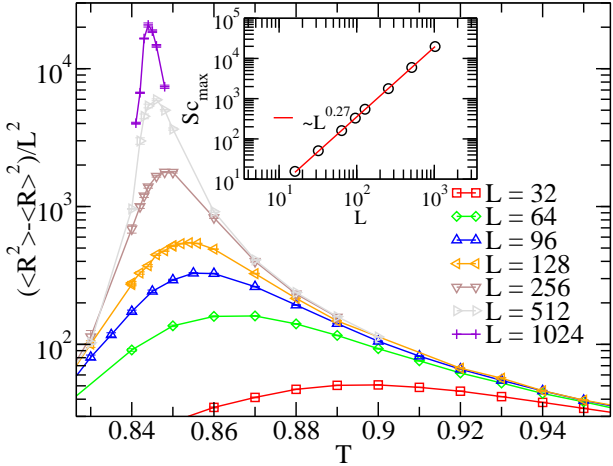}
 \caption{Nematic susceptibility of the
dimer model with only dimer-aligning interactions, $v=1$. The plaquette interaction is not included and has  $u=0$. 
The hole density is $\delta=0.03125=1/32$.
The susceptibility of the nematic order parameter displays a strong divergence with the system size (up to a lattice of $2^{20}$ sites) 
and in the inset scaling of the peak giving anomalous exponent $\eta\simeq0.25$, consistent with an Ising transition.}
 \label{fig:MC3b}
 \end{figure}

\begin{figure}[hbt]
\centering
 \includegraphics[width=0.45\textwidth]{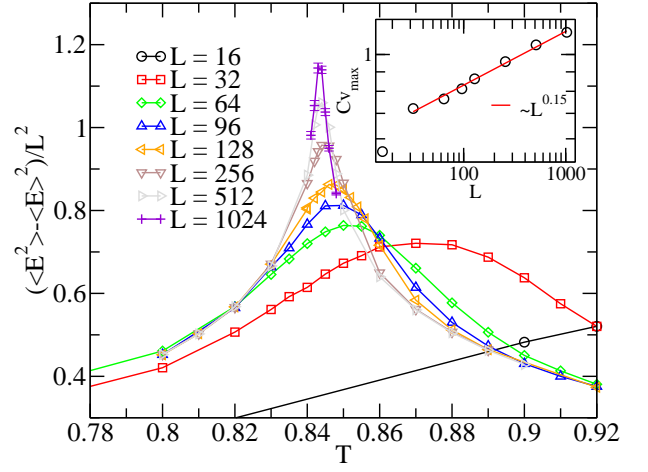}
 \caption{
 Temperature dependence of the specific heat of a
dimer model with only dimer-aligning  interactions, with $v=1$. The plaquette interaction is absent, $u=0$. 
The hole density is $\delta=0.03125=1/32$. 
The temperature dependence of the specific heat displays a clear phase transition with corresponding exponent from the peak scaling 
$\alpha/\nu\simeq0.1\pm0.1$, consistent with the expected logarithmic singularity ($\alpha=0$) at a 2D Ising transition.}
 \label{fig:MC3c}
 \end{figure}

\begin{figure}[hbt]
\centering
 \includegraphics[width=0.45\textwidth]{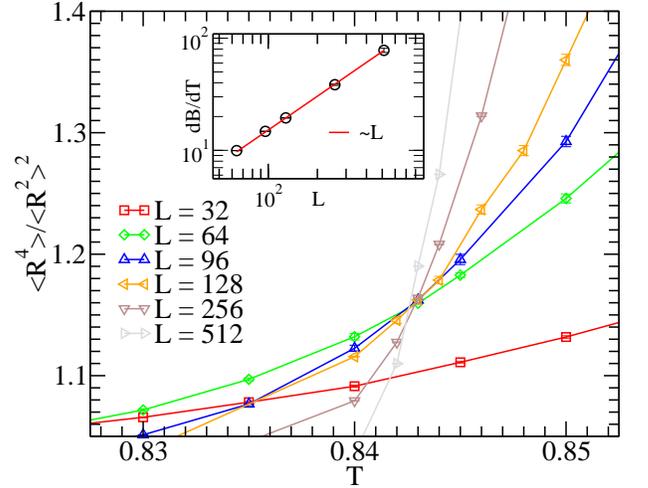}
 \caption{Binder ratio crossings for the nematic order parameter of a 
dimer model with only  dimer-aligning interactions, with $v=1$, at finite hole density. The plaquette interaction is not included, $u=0$. 
The hole density is $\delta=0.03125=1/32$. 
The observed crossing point is  consistent with the  expected 2D Ising transition and its derivative at the crossing gives clearly $\nu\simeq1$. }
 \label{fig:MC3d}
 \end{figure}

 \begin{table}[h!]
\begin{tabular}{|c|c|c|c|c|c|}
\hline 
$\delta$ & $T_c$ &$ \alpha/\nu$ & $ \eta $ &$\nu$ & $B$ \\
\hline 
0 & 0.90(2) & - & 0.25(4) & - & - \\
\hline
3.125 & 0.84(1) & 0.1(1) & 0.25(4) &1.0(1) & 1.16(5) \\
\hline
4.6875 & 0.81(1) & 0.1(1) & 0.29(6) & 1.0(1) & 1.16(4) \\
\hline
6.25 & 0.78(1) & 0.1(1) & 0.25(4) & 0.95(7) & 1.15(4) \\
\hline
9.375 & 0.72(1) & 0.1(1) & 0.26(4) & 1.1(1) & 1.13(2) \\
\hline
12.5 & 0.67(1) & 0.07(7) & 0.22(5) & 1.1(2) & 1.13(2) \\
\hline
25 & 0.49(1) & 0.03(3) &0.24(5)& 1.1(2)  &1.14(3) \\
\hline
${\rm Ising}$ & - & 0 & 1/4 & 1 & 1.13 \\
\hline
\end{tabular}
\caption{Critical behavior at the nematic-disordered phase transition line in the absence of a plaquette interaction, $u=0$. 
From Left to Right the columns denote: hole density, critical temperature, specific heat exponent ratio $\alpha/\nu$, rotational order parameter 
anomalous exponent $\eta$, correlation length exponent $\nu$ and finally, the value of the Binder cumulant at the crossing point. 
The Ising universality class prescribes the value $1.162$ for this quantity.}
\label{tab:MC3e} 
\end{table}

 \begin{figure}[t]
 \includegraphics[width=0.45\textwidth]{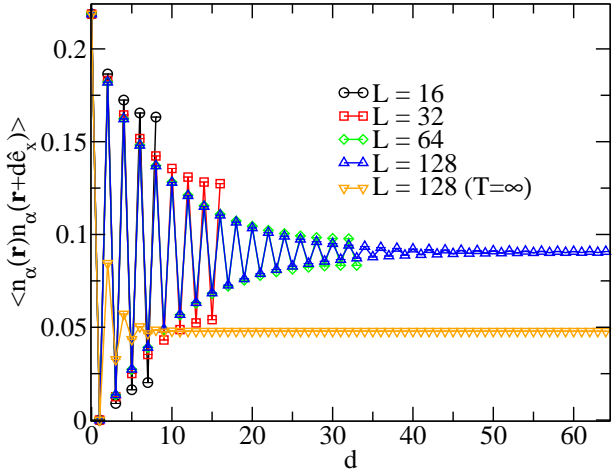}
 \caption{Finite-size dependence of the  correlations for columnar order in the nematic phase of a dimer model at finite hole density 
 with only dimer-aligning interactions.
Here, $u$=0, $T=0.5$. 
The dilution is $1/8$. The evolution of the longitudinal correlation profile is shown as a function of distance. 
Clearly, a saturation of the profile is seen and the infinite-distance value is almost twice the fully disordered value 
($T\rw\infty$, same density), consistent with the phase structure, since most dimers in the  nematic liquid are oriented in the same 
direction.  }
  \label{fig:MC3f}
 \end{figure}

\begin{figure}[t]
\includegraphics[width=0.45\textwidth]{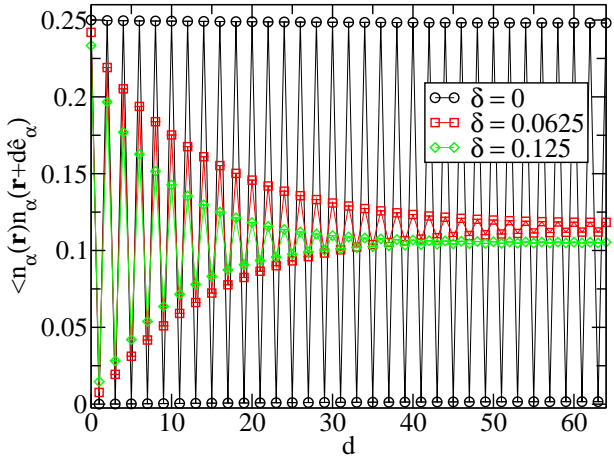}
\caption{ Columnar order correlations at different hole densities in the nematic phase for a dimer model with only dimer-aligning interactions. 
Here $L=128$ and $T=0.5$, deep in the nematic phase when at low hole densities. 
The crystal correlation at close packed dimers transforms to liquid-like at large distances immediately as dilution becomes non-zero.  }
 \label{fig:MC3g}
 \end{figure}


At finite small dilution but no plaquette interaction $u=0, v=1$, we find a clear transition line that separates the isotropic hole-dimer liquid 
phase from the Ising nematic phase. This transition is clearly in the Ising universality class, as verified by the values of the anomalous 
exponents as well as the Binder cumulant crossing values, being in consistency with the Ising universality. The evidence is presented in 
Figs.~\ref{fig:MC3a},  \ref{fig:MC3b}, \ref{fig:MC3c}, \ref{fig:MC3d} and Table~\ref{tab:MC3e}. While in the figures we show a typical case at 
dilution $\delta=1/32$ and the consistency shown by the order parameters, their susceptibilities, the specific heat and the Binder cumulants, at 
the table we show our estimations for particular critical exponents and universal cumulant values for a number of hole densities we studied, displaying a strong consistency. For $\delta=1/32$, a clear phase transition is detectable around $T = 0.9$  and the rotational symmetry is spontaneously broken at low temperature as shown by the rotational order parameter, but the translational symmetry is not, as  the vanishing, with the system size, columnar order parameter shows. The evidence for the Ising universality class comes from the specific heat which is consistently showing scaling with $\alpha\simeq0$, the rotational susceptibility which gives $\eta/\nu\simeq0.25$ and the Binder cumulant ratio which not only displays a crossing point at the expected Ising location, but also its derivative at the crossing scales with system size in a way that shows that $\nu\simeq1$. 

Finally, in Figs.~\ref{fig:MC3f}, \ref{fig:MC3g} we show how the nematic correlations evolve with system size and hole density at a low temperature $T=0.5$. As can be seen from Fig.~\ref{fig:2}C, at this temperature and all but the close-packed case, the system is an Ising nematic liquid which is stable to perturbations (such as the plaquette interaction). Through the scaling with the system size, it is clear that at long distances dimers have liquid correlations, albeit double than at infinite temperature, due to the fact that deep in the nematic phase they are primarily oriented in the same direction. The correlations have clearly converged for $L\geq64$. As a function of hole density, we observe the dramatic effect of having non-decaying correlations at $\delta=0$, signifying the absence of liquid correlations, but they immediately become liquid-like as the system gets dilute. Starting from the perfect close-packed crystal (at zero dilution, \cf Fig.~\ref{fig:2}C), assuming fixed boundary conditions, then to first order, if some dimers $n$ (introducing $2n$ holes) are removed from the lattice, then there are $\left(\begin{tabular}{c}N\\n \end{tabular}\right)$ ways to remove them, therefore contributing an extensive entropy without though destroying the columnar order. Further, to second order, at the expense of energy $1/T$, each pair of holes has a sub-extensive contribution by moving freely on the line it was created, again winning over the crystalline confining tendency. Since a pair can be formed on each line, it leads to an extensive entropy contribution as well, that can dominate over the energy cost. Therefore, the nematic liquid is also deconfining  along the oriented direction.

We should note that the Ising transition line, should end at a tricritical point and coexistence at a higher  $\delta$ from the ones we investigated. The investigation of this discontinuous transition is beyond the scope of this work, but should be in the tricritical Ising universality class.
}

\subsection{Case $u>0,v>0$, $\delta>0$: Competition between dimer-aligning  and plaquette interactions in a dimer model at finite hole density displaying  two Ising phase transitions}
\label{ssec:MC-4}

\begin{figure}[hbt]
 \includegraphics[width=0.45\textwidth]{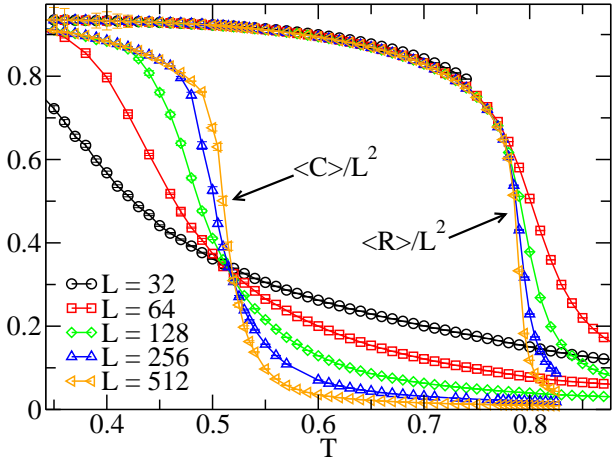}
 \caption{Evolution of the columnar and the nematic order parameters as a function of temperature in dimer model at finite hole 
 density  with both dimer-aligning, and plaquette  interactions. Here $u=0.1$ and $v=1$. 
 The  hole density is  $\delta=0.0625=1/16$. 
 Two Ising phase transitions are observed as temperature decreases. 
  The columnar order parameter shows a distinctly different transition temperature from the nematic order. 
  The isotropic-nematic transition takes place at a higher temperature due to the presence of the attractive columnar interaction. }
 \label{fig:MC4a}
 \end{figure}
 
\begin{figure}[hbt]
 \includegraphics[width=0.45\textwidth]{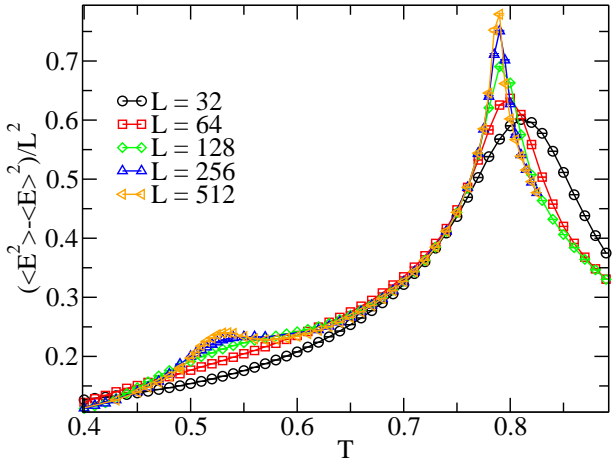}
 \caption{Specific heat for a dimer model with both dimer-aligning, and plaquette interactions at finite hole density  
 with both dimer-aligning, and plaquette  interactions. The plaquette coupling is $u=0.1$ and $v=1$. 
 The  hole density is  $\delta=0.0625=1/16$.
 The specific heat shows a signature for both transitions, but the lower temperature transition's peak is much weaker, 
 between the columnar solid and the Ising nematic liquid, converging much slower as the system size increases.}
  \label{fig:MC4b}
 \end{figure}

\begin{figure}[tb]
 \includegraphics[width=0.45\textwidth]{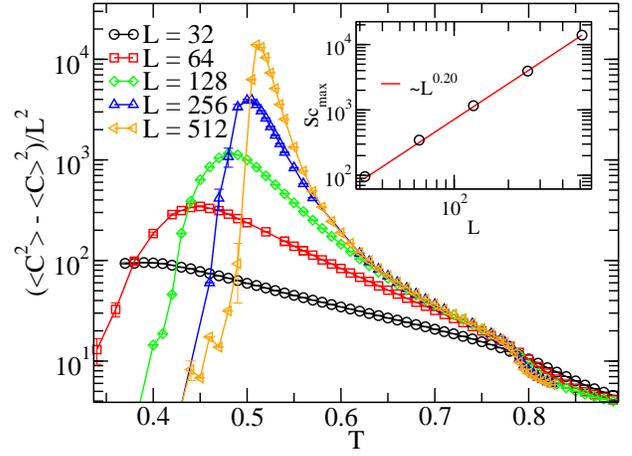}
 \caption{Columnar order susceptibility of the dimer model with both dimer-aligning and plaquette interactions at finite hole density. The plaquette coupling is  $u=0.1$ and the two (and column) coupling are $v=1$. The hole density is  $\delta=0.0625=1/16$. 
The susceptibility of the columnar order parameter displays a strong divergent peak with Ising exponents at the lower transition temperature. }
 \label{fig:MC4c}
 \end{figure}

\begin{figure}[tb]
 \includegraphics[width=0.45\textwidth]{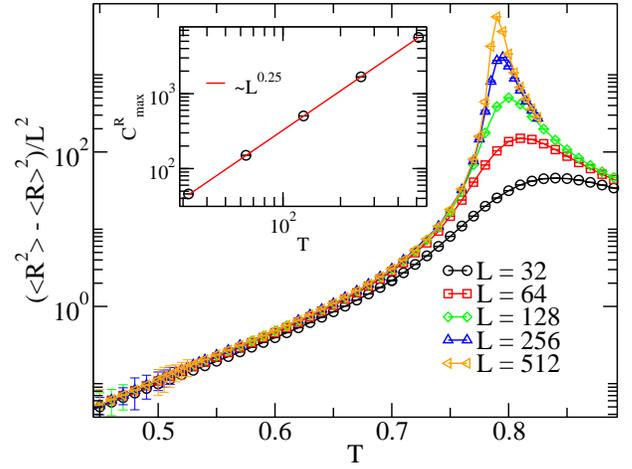}
\caption{Susceptibility for the nematic order parameter in a
dimer model with both dimer-aligning and plaquette interactions at finite hole density. The plaquette coupling is $u=0.1$ and $v=1$. The hole density is   $\delta=0.0625=1/16$. 
The susceptibility of the nematic order parameter displays a strong divergent peak with the Ising exponents at the higher transition temperature}
 \label{fig:MC4d}
 \end{figure}

\begin{figure}[tb]
 \includegraphics[width=0.45\textwidth]{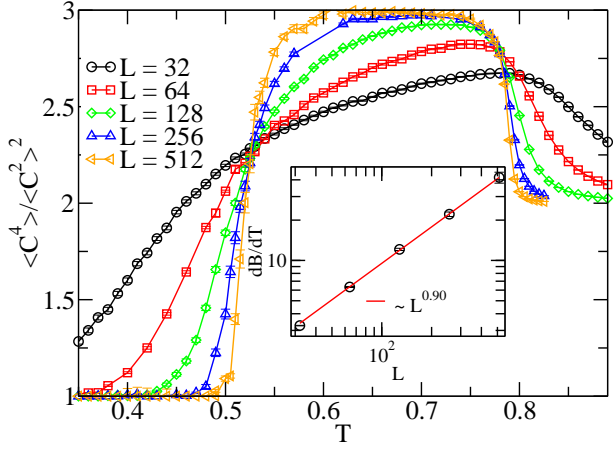}
 \caption{Binder cumulant for the columnar order parameter for a
dimer model with both dimer-aligning and plaquette interactions at finite hole density. The plaquette coupling is $u=0.1$ and the dimer-aligning couplings are $v=1$. The hole density is  $\delta=0.0625=1/16$. 
The Binder cumulant for the columnar order parameter displays crossings at both critical temperatures, since the columnar order parameter, by construction, is sensitive to both transitions in opposite ways. }
 \label{fig:MC4e}
 \end{figure}

\begin{figure}[tbh]
 \includegraphics[width=0.45\textwidth]{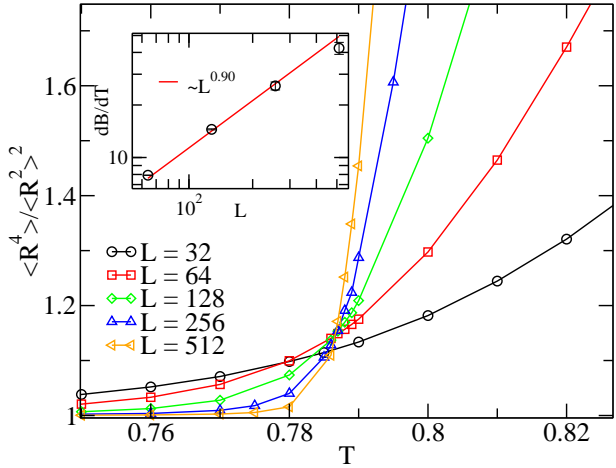}
 \caption{Binder cumulant for the nematic order parameter of a 
dimer model with both dimer-aligning  and plaquette interactions at finite hole density. The plaquette coupling is $u=0.1$ and the dimer-aligning couplings are $v=1$. The hole density is   $\delta=0.0625=1/16$. 
The Binder cumulant for the nematic order parameter displays a crossing only at the highest critical temperature with slightly lower than Ising, but consistent, critical exponent. The same overall behavior is displayed in all the hole densities studied.}
 \label{fig:MC4f}
 \end{figure}

Finally, we investigate the case of a finite hole density, where the system contains both the dimer-aligning and plaquette interactions. In this case, we observe a more complex phase diagram, where two Ising transition lines emerge from the close-packed limit (where the transition is in the KT universality class possibly, as we argued) (shown in Fig.~\ref{fig:2}B), separating the Ising nematic dimer liquid from the disordered isotropic dimer-hole liquid at high temperatures and the columnar solid phase at low temperatures. The evidence for our conclusion that both transitions are in the Ising universality class are displayed in Figs.~\ref{fig:MC4a}, \ref{fig:MC4b}, \ref{fig:MC4c}, \ref{fig:MC4d}, \ref{fig:MC4e}, \ref{fig:MC4f} for a particular hole density $\delta=1/16$ and interaction strength $u,\; v=0.1$, chosen for clarity. Our study is extended over several choices of $u$ and $\delta$ (not shown).

The columnar order parameter displays a transition at $T_C=\simeq0.5$, which is remarkably weaker than the transition displayed by the rotational order parameter at $T=T_N=\simeq0.8$. The weakness of the low temperature transition can be further observed in the behavior of the specific heat, which displays a strong peak (with a very weak divergence, as expected for Ising transitions) at $T_N$ but a very weak peak at $T_C$. 

The columnar order susceptibility displays a strong peak at $T_C$ with a peak which shows an anomalous exponent $\eta\sim0.2$, lower than the Ising-expected $1/4$, but expected given the weak convergence we observe throughout for this transition with the system size. However, the nematic order susceptibility shows a clear Ising behavior at the corresponding nematic-disordered transition, with a clear Ising anomalous exponent for over two decades of scaling, confirming the quick convergence of that transition. 

Finally, regarding the behavior of the Binder cumulants, as possibly expected, the rotational order's cumulant shows a textbook behavior at the high-temperature nematic-disordered transition, where there is a strong crossing at $~1.13$, which given our definitions, is the universal value expected for the Ising universality class. Further, the cumulant's derivative scales in a consistent way with the Ising expectations. 
However, the columnar order's cumulant displays a complex behavior with two crossings at both transitions, since the columnar order parameter displays {\it fluctuations} sensitive at both transitions. The lower transition's Binder ration displays a crossing around $\sim2.1$ that appears to drift toward lower value and  the Binder ratio derivative displays Ising-like scaling (with $\nu\sim 1$). We believe that the Binder ratio has not yet converged to the Ising value for finite-size  and finite-anisotropy reasons. Namely, the columnar order parameter is defined in a $2\times1$ unit cell while the simulation box has always been square $L\times L$. In studies of the Binder ratio for the square-lattice Ising model on rectangular boxes~\cite{selke-06}, it was shown that the critical Binder cumulant value had a strong dependence on the system's aspect ratio. We expect a strong dependence also in our case, and simulations in thin strips are expected to show a much faster convergence for this transition. 

\section{Mean-Field Theory}
\label{sec:MFT}

The understanding of this Ising nematic phase transition can be elucidated also through a mean-field study, 
similar to the approach followed in Ref.[\onlinecite{Papanikolaou:2007qf}]. 
Using a Grassmann representation for the dilute dimer model and performing two Hubbard-Stratonovich transformations, 
with the exact same field definitions, we find the following effective thermodynamic potential,
\begin{align}
\Gamma(z,V)=&\frac{V}{4}\sum_{ijkl}m_{ij}\tilde{M}_{ijkl}m_{kl} 
\nonumber\\
+&\sum_{ij}n_i\left(\frac{V}{2}\sum_{kl}\tilde{M}_{ijkl}m_{kl} + \frac{z}{2}M_{ij}\right)n_j\nonumber\\
-\sum_i \ln &\left[2\sum_{j}\left(\frac{V}{2}\sum_{kl}\tilde{M}_{ijkl}m_{kl}+\frac{z}{2}M_{ij}\right)n_j+1\right]
\end{align}
where $m_{ij},n_i$ are order parameters coupled to neighboring same-axis aligned dimers and hole density respectively, while 
$z=e^{\beta\mu}$ and $V=z^2(e^\beta-1)$ where $\beta=1/T$ and $\mu$ the hole chemical potential. 
Further, $\tilde{M}_{ijkl}$ is non-zero and unity when $i,j,k,l$ are neighboring sites on the same lattice line,
 i.e. $M_{ij}$ is non-zero and unity when $ij$ are nearest neighboring sites. The only difference with the study of 
 Ref.[\onlinecite{Papanikolaou:2007qf}] is that the attractive neighboring interaction is in a different direction (aligning instead of plaquette). 

\begin{figure}[hbt]
 \centering
 \includegraphics[width=0.35\textwidth]{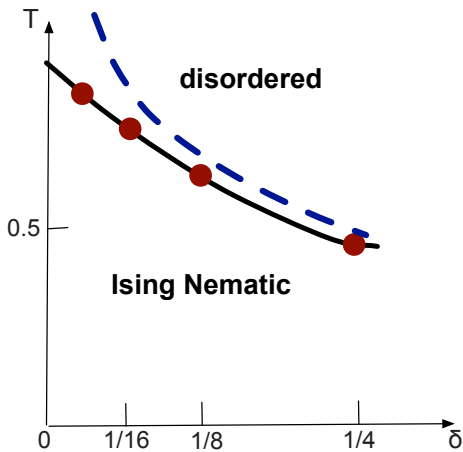}
  \caption{
  Mean-Field Approach to the Phase diagram in the $u=0, \; v=1$ case. 
  The thick dashed line describes a mean-field estimate from an associated mean-field theory discussed in the text. The points signify the transition line, as estimated from numerical simulations.}
 \label{fig:mf}
 \end{figure}

We use the ansatz $n_{i}=n$ and $m_{ij}=(-1)^im$ and we look for solutions of the extremal equation
\begin{equation}
\frac{\partial \Gamma}{\partial m}=0, \qquad \textrm{with} \quad \beta z \frac{\partial \Gamma}{\partial  z}=n
\end{equation}
 The equations can be either solved numerically or analytically, with the latter being in an expansion in $z$ (small dimer density) and 
 then, $\beta$ (high temperature), giving finally the leading term for the critical hole density at high temperatures  to be 
 \begin{equation}
 \delta_c(T)=2(1+\beta-\sqrt{1+2\beta})/\beta=\beta+O(\beta^2)
 \end{equation}
 leading to the dashed line shown in Fig.~\ref{fig:mf}.

\section{Conclusions}
\label{sec:conclusions}

In this paper we examined the phase diagram of a 2D dimer model on the square lattice at zero and finite hole densities. 
The important difference between this work and previous studies is the consideration of the interplay of attractive  interactions 
on a plaquette (which was considered previously) with attractive (aligning) dimer interactions on the same row (and column) 
for dimers on next-nearest neighboring bonds of the lattice. By means of large-scale Monte Carlo simulations, combined 
with a scaling analysis and a mean-field theory we find that the phase diagram of this simple model, shown in Fig. \ref{fig:2}, 
in addition to a line of fixed points at zero hole density,  generally contains three phases: a dimer-hole liquid, a columnar phase 
(valence bond solid) and a novel Ising nematic phase. We presented detailed results for the scaling behavior of the columnar 
and nematic (orientational) order parameters, specific heat and Binder cumulants at the various transitions. 
In the general case our results are consistent with two Ising transitions as the interaction and hole density are varied.

These results are of interest to both the understanding of classical dimer models and to quantum dimer models which arise 
in the context of frustrated quantum magnets. In the latter context the theory we presented here describes the behavior of 
RVB-type wave functions at finite hole density. All phases described here are non-magnetic as the dimers represent valence bond singlets.

Experimentally relevant models of nematic phases should have a columnar phase dominating at low temperatures, 
consistent with strong experimental evidence in cuprate and iron pnictides materials. In this work, we considered a 
framework that includes geometric frustration originating in the hard-core dimer exclusion effect on the square lattice. 
We find that exactly because of the geometric frustration, a nematic phase at low temperatures is {\it always} unstable 
to columnar fluctuations, due to entropic reasons, implying that negligible amounts of the plaquette interaction $u$ 
yields a columnar phase at low temperatures. It would be very interesting to investigate similar phase diagrams on 
different lattices such as the kagome, where a $Z_2$ spin-liquid has been identified as the ground-state of the Heisenberg 
antiferromagnet\cite{Yan:2011uq} pointing toward useful dimer descriptions 
for the ground-state wavefunction.\cite{Schwandt:2010fk,Rousochatzakis:2013kx}

\begin{acknowledgments}
We would like to thank Cristian Batista, Steven Kivelson, J{\"o}rg Schmalian and Bernard Nienhuis for inspiring discussions.
This work is supported in part by the National Science Foundation through the grant DMR-1064319 at the University of Illinois (EF), 
and through a DTRA grant No. 1-10-0021 (SP) at Yale University. We note that we used the ALPS libraries~\cite{alps}.
This work benefited greatly from the facilities and staff of the Yale University Faculty of Arts and Sciences High Performance 
Computing Center as well as the computing facilities of MPI-PKS Dresden 
\end{acknowledgments}

\bibliographystyle{apsrev}


\end{document}